# Robust Deep Learning Framework for Constitutive Relation Modeling

Qing-Jie Li[a,*], Mahmut Nedim Cinbiz[b], Yin Zhang[a], Qi He[a], Geoffrey Beausoleil II[b], Ju Li[a,c,*]

[a] Department of Nuclear Science and Engineering, Massachusetts Institute of Technology, Cambridge, MA, 02139, USA

[b] Idaho National Laboratory, Idaho Falls, ID, 83415, USA

[c] Department of Materials Science and Engineering, Massachusetts Institute of Technology, Cambridge, MA, 02139, USA

Correspondence: qjl@mit.edu, liju@mit.edu

Modeling the full-range deformation behaviors of materials under complex loading and materials conditions is a significant challenge for constitutive relations (CRs) modeling. We propose a general encoder-decoder deep learning framework that can model high-dimensional stress-strain data and complex loading histories with robustness and universal capability. The framework employs an encoder to project high-dimensional input information (e.g., loading history, loading conditions, and materials information) to a lower-dimensional hidden space and a decoder to map the hidden representation to the stress of interest. We evaluated various encoder architectures, including gated recurrent unit (GRU), GRU with attention, temporal convolutional network (TCN), and the Transformer encoder, on two complex stress-strain datasets that were designed to include a wide range of complex loading histories and loading conditions. All architectures achieved excellent test results with an RMSE below 1 MPa. Additionally, we analyzed the capability of the different architectures to make predictions on out-of-domain applications, with an uncertainty estimation based on deep ensembles. The proposed approach provides a robust alternative to empirical/semi-empirical models for CRs modeling, offering the potential for more accurate and efficient materials design and optimization.

## 1. Introduction

An mm-sized representative volume element (RVE) in an engineering continuum mechanics model can physically contain ~$10^{20}$ atoms, ~$10^{14}$ point defects, ~$10^5$ m of mobile dislocations, numerous grain boundaries, hetero-phases, cracks, etc. It is clearly impossible to treat these degrees of freedoms (DOFs) on-the-fly for modeling and designing metal stamping, extrusion, crash-worthiness, etc., at the component scale. So the burden of coarse-graining over all these DOFs, which could certainly evolve under stress in a history-dependent manner, falls onto the so-called constitutive relations (CRs), as far as the macroscopic responses are concerned. The development of reliable CRs is critical for using computational mechanics to support component qualifications in automotive, aerospace and nuclear applications. Over the past many decades, various empirical/semi-empirical constitutive models have been developed. Based on the large strain behavior, plastic constitutive relations can be categorized as Voce type, Holomon type, or their combinations [1]. While Voce type relations tend to saturate [2–8], Holomon (power law) type of constitutive relations are unbounded at large strains [9,10]. These physically and phenomenologically based models have been widely used for different materials. However, one major limitation associated with these models is the insufficient capability to simultaneously describe various deformation stages under different conditions [11]. For example, most models only focus on specific stages of the entire deformation process and can only deal with a few external conditions (e.g., temperature and strain rate). Depending on the scenarios, many separate/modified models need to be constructed, such as in modeling various hardening behaviors [12–17]. For materials experiencing complex dynamic evolutions or materials optimization in a high-dimensional parameter space, a robust constitutive modeling calls for universal model (e.g., structural materials in nuclear applications can be simultaneously subjected to compositional/structural/environmental changes).

Artificial neural networks (NN) have long been considered universal approximators [18,19], which naturally can serve as a robust alternative to empirical/semi-empirical constitutive modeling. The application of NN in constitutive modeling is not a new idea; some inspiring attempts were made nearly two to three decades ago. For example, Ghaboussi et al. [20] first introduced NN to model the mechanical behavior of plain concrete that was subjected to both biaxial loading and uniaxial cyclic loading. Ghaboussi et al. also proposed an auto-progressive algorithm for training NN to learn complex constitutive relations from global load-deflection response [21]. Lefik et al. implemented an incremental NN representation of constitutive relations that was successfully incorporated into a Finite Element (FE) code [22]. Hashash et al. [23] derived a consistent material stiffness matrix to address the numerical implementation issues for NN-based FE analysis. Jung and Ghaboussi[24] developed a rate-dependent NN visco-elasticity constitutive model and implemented it in FE analysis.

The dramatic boost in computational power and breakthroughs in newer NN architectures have led to a resurgence in constitutive modeling. Based on the inputs, three types of approaches can be identified for NN-based constitutive learning:

1. In the first approach, constitutive relations are learned from direct stress-strain responses. Such stress-strain data can be obtained from uniaxial or multi-axial loadings and include loading histories. For example, Gorji et al. developed a recurrent NN (RNN)-based framework for modeling the large deformation response of elasto-plastic solids subjected to arbitrary multi-axial loading paths [25].

2. Instead of learning constitutive relations from direct stress-strain data, the second approach learns the underlying constitutive relations from indirect measurements such as full field load-displacement data. For instance, Xu et al. [26] proposed an inverse modeling scheme to learn the constitutive relations of viscoelastic materials from indirect displacement data. Zhang et al. [27] developed a hybrid FE-NN framework to learn constitutive relations from full-field displacement data. Such indirect constitutive learning usually needs to consider physical constraints that are often in the form of partial differential equations.

3. The third approach takes advantage of prior knowledge, principles, or empirical models. For example, Li et al. [28] replaced the strain rate and temperature terms in Johnson-Cook model with a NN to describe the non-monotonic temperature dependence. Linka et al. [29] developed a CANNs (constitutive artificial neural networks) framework that learns a generalized strain energy function to link strain and materials information to stress responses. Incorporating prior knowledge or principles of mechanics and materials theory may reduce training data and achieve better extrapolation.

The mechanical response of materials is commonly characterized using simple, standardized uniaxial tensile test articles. The basic mechanical properties (elastic modulus, yield stress, ultimate stress, rupture strain, and ductility) and plastic constitutive relations (such as temperature sensitivity, strain and strain-rate hardening coefficients) are deduced from one-dimensional stress-strain curves. As the macroscopic strain-stress response is an effective and widely used measure of material deformation behaviors for engineering design, extensive strain-stress data sets are being accumulated from past and present research efforts. These data sets are also becoming more complicated as additional manufacturing methods yield additional features for consideration, e.g., through 3D printing and high throughput testing [30]. It is therefore beneficial to take advantage of strain-stress data to build robust constitutive models. In this work, we propose a general encoder-decoder deep learning framework to model complex high-dimensional stress-strain data. In the proposed framework, an encoder first projects the complex high-dimensional input data, such as loading histories, loading conditions, and materials information, onto a lower-dimensional hidden space. Such hidden representation of input information is then mapped to stresses of interest via a decoder. We evaluated a series of encoder architectures that are capable of modeling temporal data, such as gated recurrent unit (GRU) [31], GRU with attention, temporal convolutional network (TCN) [32], and transformer encoder[33]. The decoder was implemented as a fully connected network (FCN); however, it can be replaced with sequence modeling architectures if a series of stress prediction is needed. We tested all encoder architectures on two complex datasets: 1) a one-dimensional stress-strain curve with multiple unloading-reloading cycles to include complex loading history information; 2) a synthetic dataset based on the Johnson-Cook model [34] to include complex loading conditions such as temperature and strain rate. All architectures demonstrate excellent test results, with a root mean squared error (RMSE) well below 1 MPa, which is often needed to satisfactorily capture the full-range deformation features. We also explored the capability of different architectures on out-of-domain predictions with uncertainty estimation based on an ensemble of models[35]. Overall, our proposed general deep learning framework demonstrates high accuracy and universal capability for modeling various challenging CRs, thus offering a robust alternative to conventional CRs modeling for materials design and optimization.

## 2. Methods

**2.1 Learning task and general encoder-decoder deep learning framework**

From a fundamental point of view, the stress-strain relation of a material stems from the collective interatomic (and/or intermolecular) responses upon external loading, which involves various atomistic processes on different time-/length-scales. While the objective of our deep learning approach is to accurately and efficiently predict stress-strain relations in a physics-agnostic way, understanding the underlying deformation processes and material features on various spatial and temporal scales can be helpful in preparing meaningful training data. For example, to effectively model the macroscopic stress-strain relations, we may consider time-related data such as loading rate and temperature (as in thermally activated processes), and measurable material information on relevant lengthscale (e.g., chemical composition, crystal structure, defect population density). We can denote the measurable materials information using a vector **m** that may include composition, grain size, phase fractions, dislocation density, etc. Then, our learning task can be defined as

$$\boldsymbol{\sigma} = NN([\mathbf{m}, \boldsymbol{\varepsilon}, \mathbf{t}]) \tag{1}$$

where $NN$ means the NN model to be learned, $\boldsymbol{\sigma}$ is the stress, **m** is the measurable material information (note that **m** represents only the measurable part of the material microstructure, as much of the microstructural details are hidden from view and not measured), $\boldsymbol{\varepsilon}$ represents loading history, and **t** includes time-scale information such as strain rate, temperature, corrosion rate, radiation exposure, etc.

In the context of history-dependent NN modeling, the input information in Eqn. (1) is usually rewritten as a time sequence:

$$\mathbf{s} = \begin{bmatrix} \mathbf{x}_{t-n} \\ \ldots \\ \mathbf{x}_{t-2} \\ \mathbf{x}_{t-1} \\ \mathbf{x}_t \end{bmatrix} \tag{2}$$

where the component $\mathbf{x}_{t-i}$ in **s** is a vector containing input information at time $t-i$,

$$\mathbf{x}_{t-i} = [m_{t-i}^1, \ldots, m_{t-i}^{N1}, \varepsilon_{t-i}^1, \ldots, \varepsilon_{t-i}^6, \dot{\varepsilon}_{t-i}^1, \ldots, \dot{\varepsilon}_{t-i}^1, T_{t-i}] \qquad (3)$$

Such input at time $t-i$ contains $N_1$ measurable materials information (although a subscript *t-i* is used, they are just constants serving to distinguish materials), 6 strain components, 6 strain-rate components, and temperature. Depending on specific scenarios, Eqn. (3) can be reduced to fewer components. For example, in the case of a single material (no need to distinguish from others) subjected to uniaxial tensile loading, different strain rates, and a range of temperatures, Eqn. (3) can be simplified as $\mathbf{x}_{t-i} = [\varepsilon_{t-i}^1, \dot{\varepsilon}_{t-i}^1, T_{t-i}]$. Or if one wants to model a single uniaxial stress-strain relation (i.e., a single material at a fixed strain rate and temperature), then Eqn. (3) is further reduced to $\mathbf{x}_{t-i} = [\varepsilon_{t-i}^1]$. The output of a NN model (Eqn. (1)), in general, is a vector $\boldsymbol{\sigma} = [\sigma_1, \ldots, \sigma_6]$ that contains 6 stress components. However, $\boldsymbol{\sigma}$ can also be reduced to less components in cases such as uniaxial loading or two-dimensional loading.

To tackle the above outlined learning task, we propose a deep learning framework consisting of an encoder and a decoder (Fig. 1). The framework maps the high-dimensional input sequence data **s** to a lower-dimensional space and then learns to predict stresses of interest based on the encoded representation. The encoder plays a critical role in the framework by 1) capturing temporal correlations in the loading histories, and 2) effectively reducing the dimensionality of the input sequence data while preserving the most important information. Several popular architectures for handling sequence or temporal data, such as long short term memory (LSTM)[36], gated recurrent unit (GRU)[31], temporal convolutional network (TCN)[32], and transformer[33], can be used as the encoder. These architectures have already achieved impressive results in natural language processing and time series forecasting. By explicitly imposing the temporal constraint in these encoder architectures, the model training will be guided to more physically relevant regions in the parameter space, thus potentially improving its extrapolation capability. For the decoder, we can use a fully connected network (FCN) if the output is a single scaler stress, or relevant sequence modeling architectures if the output is a series of stresses. The availability of various encoder/decoder architectures provides a rich set of tools to choose from and enables the framework to adapt to different types of data and modeling objectives. In the following, we briefly introduce several encoder architectures relevant to this work.

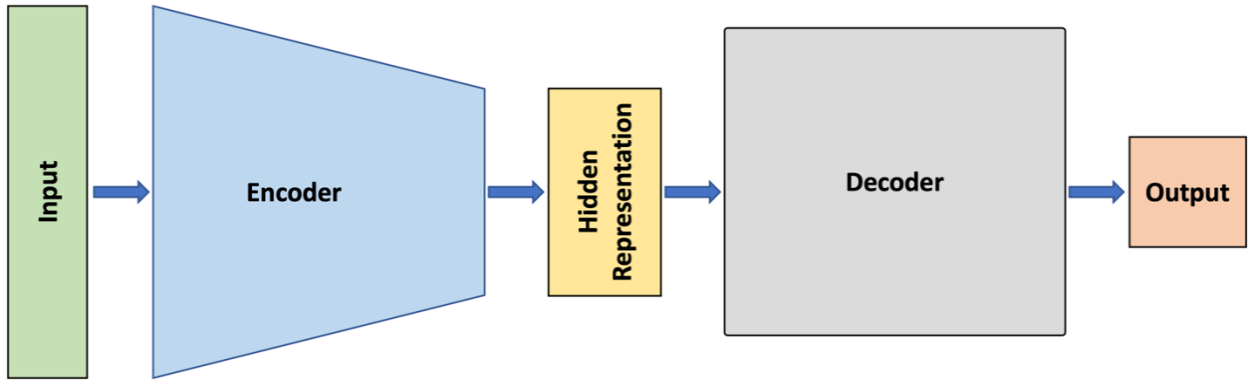

*Fig. 1 The general encoder-decoder deep learning framework for modeling constitutive relations. The input is usually a high-dimensional sequence data that contain loading histories, loading conditions, and materials information. The encoder projects the high-dimensional input information onto a lower-dimensional hidden space, respecting the temporal correlations. The hidden representation is then mapped to stresses of interest via a decoder. The decoder can be a fully connected network or sequence modeling architectures depending on the need.*

**2.2 Gated recurrent unit-based framework**

Gated recurrent unit (GRU)[31], a variation of the LSTM network, is a type of recurrent neural network (RNN) that can effectively retain long-term dependencies in sequential data such as speech and text, and has been widely adopted for natural language processing tasks, as well as sensing and guiding dynamical actions with policy NN in gaming and robotics. Here we use GRU to process loading history in an RVE. As shown in Fig. 2(a), in a GRU-based RNN, a GRU cell takes 1) the hidden state $\mathbf{h}_{t-1}$ from previous time *t*-1, and 2) current input $\mathbf{x}_t$ as inputs to update the current hidden state

$$\mathbf{h}_t = f(\mathbf{h}_{t-1}, \mathbf{x}_t) \qquad (4)$$

where *f* represents the GRU nonlinear activation function. A more compact representation of the GRU-based RNN is shown in Fig. 2(b). GRU uses gating mechanisms to control information flow between recurrent cells. As shown in Fig. 2(c), to compute the hidden state $\mathbf{h}_t$, we first compute a *reset gate* $\mathbf{r}_t$ and an *update gate* $\mathbf{z}_t$, separately, according to

$$\mathbf{r}_t = \text{sigma}(\mathbf{W}_r \mathbf{x}_t + \mathbf{U}_r \mathbf{h}_{t-1} + \mathbf{b}_r) \quad (5)$$

$$\mathbf{z}_t = \text{sigma}(\mathbf{W}_z \mathbf{x}_t + \mathbf{U}_z \mathbf{h}_{t-1} + \mathbf{b}_z) \quad (6)$$

where $\mathbf{W}$, $\mathbf{U}$, and $\mathbf{b}$ are weight/bias matrices/vectors to be learned (subscripts r and z correspond to reset and update gate, respectively), and sigma(·) is the sigmoid activation function. Then the reset gate $\mathbf{r}_t$ is further used to compute a *candidate hidden state* $\tilde{\mathbf{h}}_t$

$$\tilde{\mathbf{h}}_t = \tanh\left(\widetilde{\mathbf{W}} \mathbf{x}_t + \tilde{\mathbf{b}}_x + \mathbf{r}_t \odot \left(\widetilde{\mathbf{U}} \mathbf{h}_{t-1} + \tilde{\mathbf{b}}_h\right)\right) \quad (7)$$

where $\widetilde{\mathbf{W}}$, $\widetilde{\mathbf{U}}$, and $\tilde{\mathbf{b}}$ are learnable weight/bias matrices/vector and tanh is the hyperbolic tangent activation function. Finally, the hidden state at the current time *t* is updated as

$$\mathbf{h}_t = \mathbf{z}_t \mathbf{h}_{t-1} + (1 - \mathbf{z}_t)\tilde{\mathbf{h}}_t \quad (8)$$

If the reset gate $\mathbf{r}_t \sim 0$, the information from previous hidden states will be dropped out and $\tilde{\mathbf{h}}_t$ will be reset with current input information, while the update gate $\mathbf{z}_t$ determines how much information from the previous hidden state should be retained. Together, such gating mechanisms can capture the most relevant information from a long sequence of data stream for predicting the quantity of interest. Operations associated with Eqn. (5)-(8) are outlined in Fig. 2(c) using dashed rectangles. One may even stack multiple GRU-based RNNs to form a multilayer architecture, as shown in Fig. 2(d). In this case, hidden states for the first GRU layer are updated using the same procedure as above. For an intermediate layer *i*, the hidden state at time *t*, $\mathbf{h}_t^i$, is updated according to

$$\mathbf{h}_t^i = f\left(\mathbf{h}_{t-1}^i, \mathbf{h}_t^{i-1}\right) \quad (9)$$

where $\mathbf{h}^i_{t-1}$ is the previous hidden states in layer $i$ and $\mathbf{h}^{i-1}_t$ is the current hidden state in previous layer $i$-1. Such multilayer GRU-based RNN allows further abstraction of the raw sequence data.

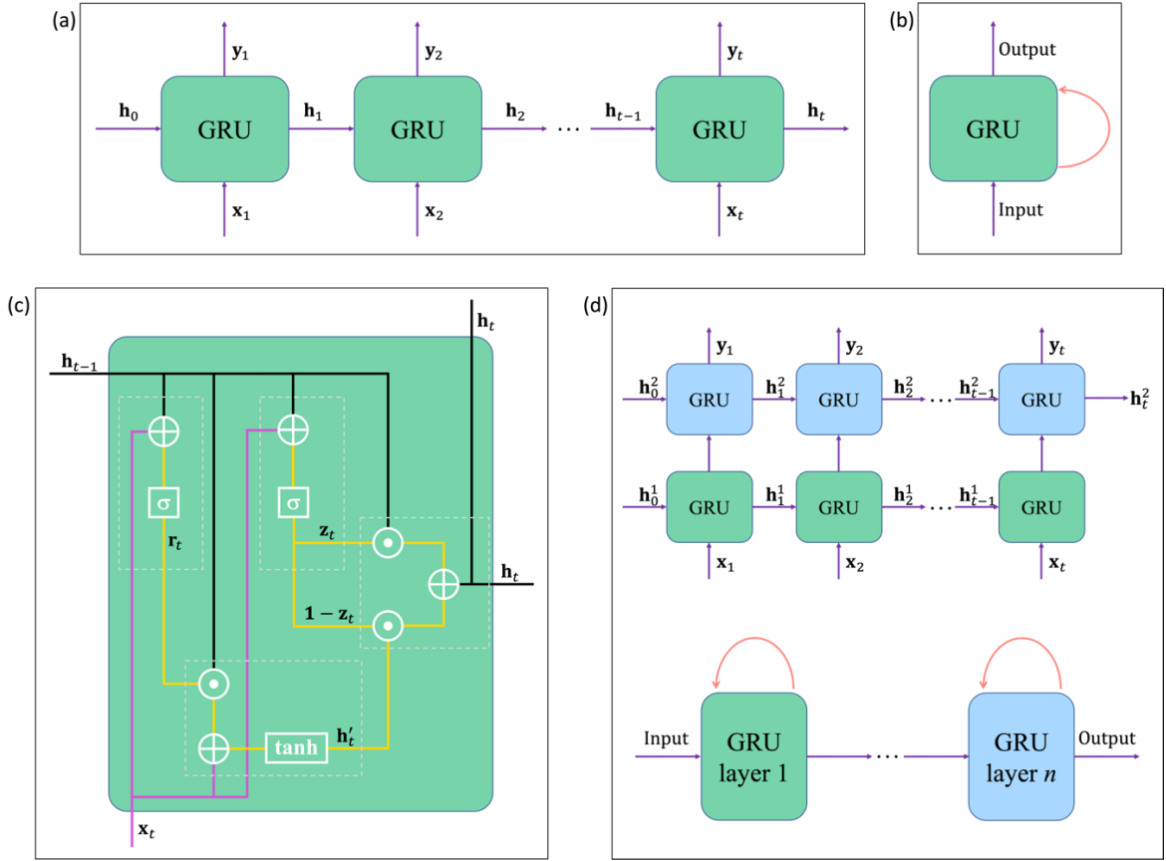

*Fig. 2 Schematic illustration for GRU-based RNN architecture. (a) RNN consisting of GRU cells. The hidden state $\mathbf{h}_t$ at time t is updated based on the hidden state $\mathbf{h}_{t-1}$ and current input $\mathbf{x}_t$; an optional output $\mathbf{y}_t$ can also be generated. (b) A convenient representation for GRU-based RNN. (c) The gating mechanisms underlying a GRU cell. The information flow of hidden states and the input at time t are rendered in black and magenta, respectively, while the intermediate information flow is colored in yellow. Plus sign, sigma sign, circled dot, and tanh sign represents element-wise summation, sigmoid activation, Hadamard product, and tanh activation, respectively. Components associated with the reset gate $\mathbf{r}_t$, update gate $\mathbf{z}_t$, tentative new hidden states $\mathbf{h}'_t$, and updated hidden state $\mathbf{h}_t$ are outlined with dashed lines rectangles, respectively. (d) An example of RNN consisting of two GRU layers (upper panel) and a convenient representation of multilayer GRU RNN.*

Let us take uniaxial loading as an example to demonstrate how to fit GRU into the encoder-decoder framework. To predict the current stress $\sigma_t$ for a given loading history vector $\boldsymbol{\varepsilon} = \{\varepsilon_0, \varepsilon_1, \ldots, \varepsilon_{t-1}, \varepsilon_t\}$, we first extract the final hidden state vector $\mathbf{h}^n_t$ from a $n$-layer GRU encoder.

Then $\mathbf{h}_t^n$ is passed to an FCN that further outputs the predicted stress $\hat{\sigma}_t$. The entire architecture is schematically shown in Fig. 3. Gorji et al.[25] successfully applied a similar architecture to predict the stress-strain responses for materials.

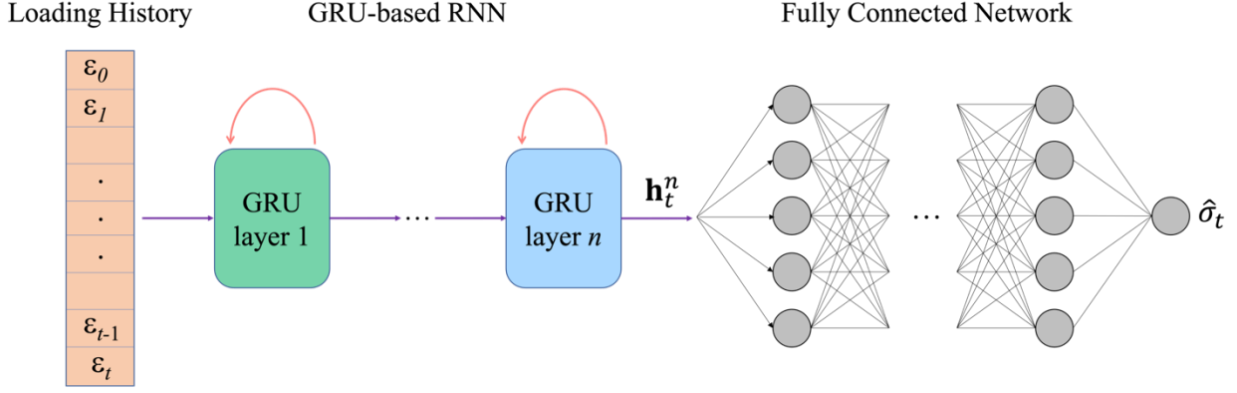

*Fig. 3 GRU-encoder based architecture to model stress-strain responses.*

## 2.3 GRU-attention based framework

The GRU-encoder architecture described above only uses the final hidden states $\mathbf{h}_t^n$ from the last GRU layer, and discards all the previous hidden states in the same layer. However, all these previous hidden states contain valuable historical information and are easily accessible from computer memory. By properly retrieving information from these stored hidden states, we can obtain useful contextual information. One effective way to achieve this is by using an attention mechanism[37]. This mechanism allows the decoder to automatically focus on the most relevant parts of the previous hidden states. The scaled dot-product attention mechanism[33] has been shown effective in different tasks:

$$\text{Attention}(Q, K, V) = \text{softmax}\left(\frac{QK^T}{\sqrt{d_k}}\right)V \quad (10)$$

where $Q$ is a query, $K$ is the keys interacting with the query, $d_k$ is the dimension of the query vector and key vectors, $V$ is the values to be summed with a weight factor from the softmax function. In our case, the final hidden state $\mathbf{h}_t^n$ serves as a query $Q$ which interacts (via dot-product) with all previous hidden states (keys $K$, which are obtained through a linear transformation of previous hidden states), to determine how much information (softmax function) from each previous hidden state should be retrieved (values $V$, which are obtained via another linear transformation of previous hidden states). Such attention layer outputs a context vector $\boldsymbol{c}_t$, which is then concatenated with $\mathbf{h}_t^n$ to form the hidden representation of input sequence data. This hidden representation is then mapped to stresses through a FCN decoder. The overall architecture is shown in Fig. 4.

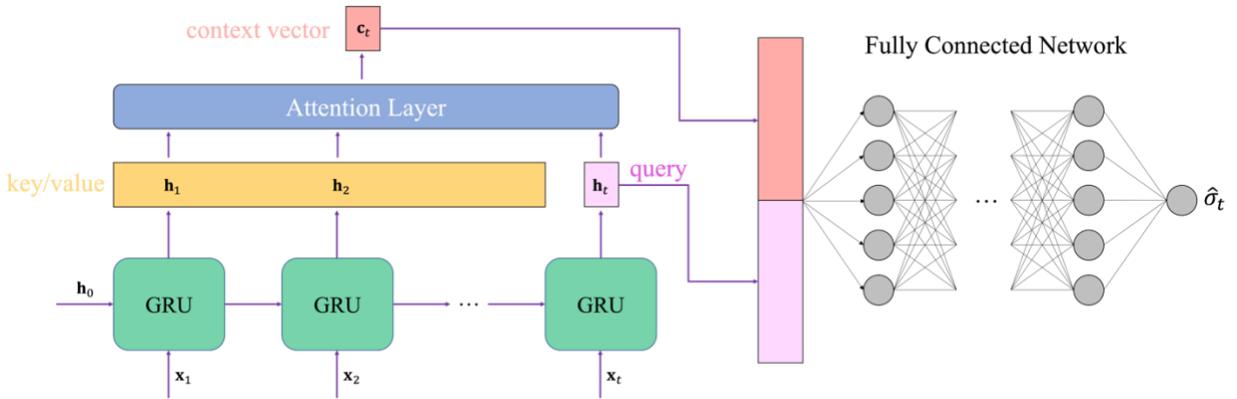

*Fig. 4 GRU-attention based architecture for modeling stress-strain data.*

## 2.4 Temporal convolutional network (TCN) based framework

Temporal convolutional network TCN [32] has been shown as a robust method for various sequence modeling tasks [32]. The original TCN architecture tries to predict an output sequence $y_0, \ldots, y_T$, given an input sequence $x_0, \ldots, x_T$:

$$\hat{y}_0, \ldots, \hat{y}_T = f(x_0, \ldots, x_T) \qquad (11)$$

Such mapping must satisfy a causal constraint that each of the component $y_t$ in the output sequence only depends on $x_0, \ldots, x_t$, meaning that there is no information "leakage" from the future to the past. Usually the causality constraint is achieved using *causal convolutions* in a 1D fully convolutional network, Fig. 5(a), i.e., the output at time *t* is convolved only with elements up to *t* in the previous layer. For example, in Fig. 5(a), the last element (magenta) in the first hidden layer is convolved only with $x_{T-2}, x_{T-1}, x_T$ if the filter size $k=3$. However, such simple causal convolution may require very deep network to span a sufficiently long history (i.e., the receptive field). To solve this problem, dilated convolutions are usually employed to effectively increase the receptive field. For example, if a dilation factor $d = 2$ is used, as shown in Fig. 5(a) from the first hidden layer to the second hidden layer, a fixed step of 2 will be introduced between two adjacent filter taps. A common practice is to exponentially increase the dilation factor $d$ with the depth of the network. Overall, the receptive field of TCN depends on the network depth, filter size $k$, and dilation factor $d$. For very large history size (e.g., $10^3$), even if we employ dilation convolution with a relatively large filter size $k$, it may still require a very deep network, which can lead to serious stabilization issues of TCN. Therefore, the convolutional layer, as shown in Fig. 5(a), is often replaced by a residual block (Fig. 5(b)) that learns the residual to the identity mapping of input, which benefits deep network learning tasks. In Fig. 5(b), the residual block consists of two dilated causal convolutional layers, each of them followed by weight normalization, the ReLU nonlinear activation, and spatial dropout for regularization. See Ref. [32] for more details. We note that Ghaboussi and Sidarta[38] in 1998 proposed a nested adaptive neural network (NANN) architecture to account for loading history dependence. In NANN, more distant loading history only has a one-way connection to more recent loading history, which also prevents information leakage from the future.

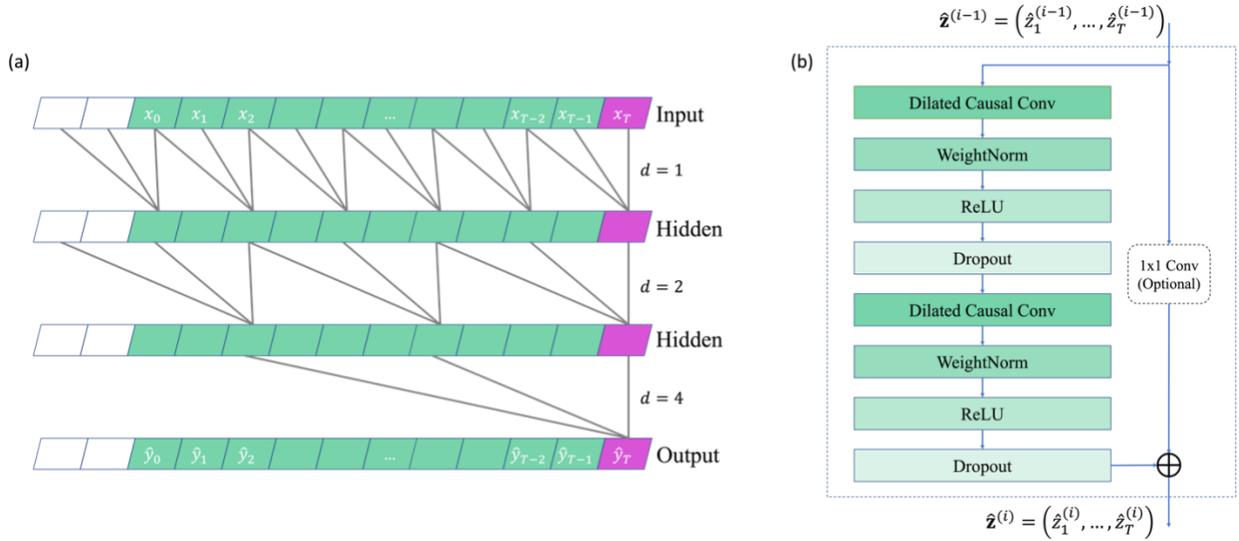

*Fig. 5 Architecture of temporal convolutional network. (a) Causal and dilated convolutions in a 1D convolutional network. Zero paddings (white cells) are applied to ensure the same length for each convolutional layer. (b) Residual block as employed in the TCN architecture. The residual block consists of two dilated causal convolutional layers, each of them followed by weight normalization, ReLU activation, and spatial dropout operations. A skip connection is made from the input to the output of the second Dropout operation. A 1×1 convolution will be used if the residual input and output have different dimensions. More details can be found in Ref. [32].*

Fig. 6 shows the overall encoder-decoder architecture using TCN as an encoder. As seen in the case of a uniaxial loading, the TCN encoder takes a loading history vector $\boldsymbol{\varepsilon} = (\varepsilon_0, \varepsilon_1, \ldots, \varepsilon_t)$ as input and outputs the hidden states associated with each input element, $\mathbf{h} = (\mathbf{h}_0, \mathbf{h}_1, \ldots, \mathbf{h}_t)$. Then the hidden state $\mathbf{h}_t$ corresponding to $\varepsilon_t$ is chosen as hidden representation of the input sequence which is passed to an FCN decoder to predict stress $\hat{\sigma}_t$. Note that one can use multiple channels/filters for the causal dilated convolutional layers in TCN, thus $\mathbf{h}_t$ in general is a vector that contains the most relevant loading history information to predict the current stress.

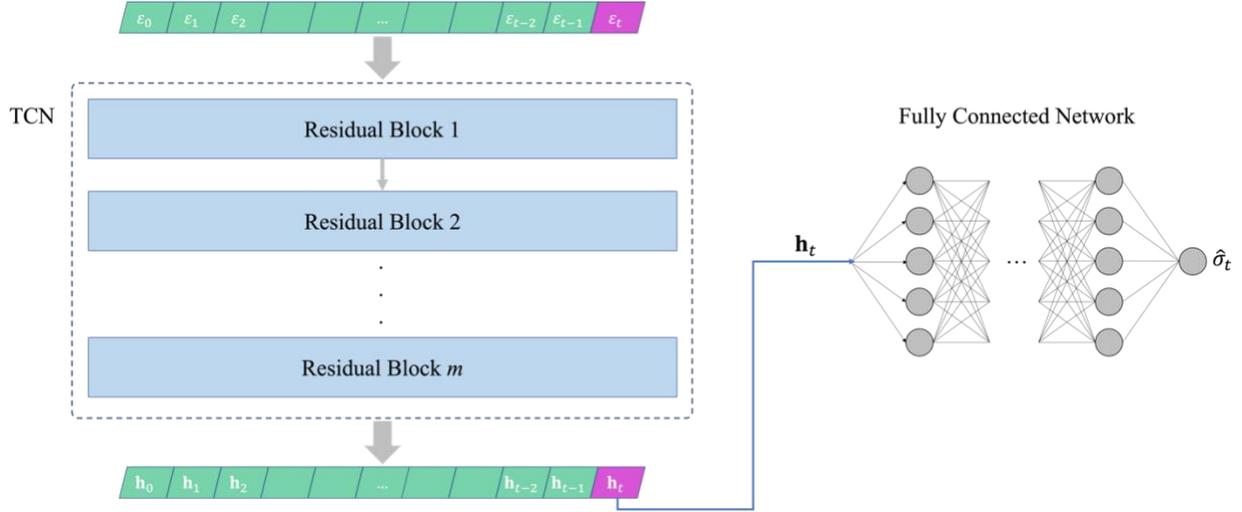

*Fig. 6 Temporal convolutional network based architecture to model strain-stress relation. The TCN takes a loading history consisting of a strain sequence as input and output the hidden states at each time. The hidden state at time t, $\mathbf{h}_t$, is then passed to a fully connected feed-forward network to predict the stress at time t, $\hat{\sigma}_t$. Multiple channels/filters can be used in each of the residual block of TCN, thus the hidden state $\mathbf{h}_t$ is a vector in general.*

**2.5 Transfomer encoder based framework**

The Transformer architecture [33] has become a key component of natural language processing and is widely used in applications such as machine translation, sentiment analysis, and text generation. Unlike recurrent neural networks that process sequences sequentially, the Transformer architecture uses a novel self-attention mechanism that allows parallel processing of sequence. This achieves faster training and inference, as well as better performance on long-range dependences in sequence modeling tasks. In this work, we utilize the Transformer encoder to learn the hidden representation of input information.

As illustrated in Fig. 7, our framework first projects all components in the input sequences to a higher-dimensional space using a linear layer. Then, a positional encoder adds positional information to the linearly transformed input. The encoded input sequence is then fed into the Transformer encoder layer, which consists of a masked attention layer and a feed forward network. The masked attention layer only allows a query to attend to all previous components, thereby preventing information leakage from future components. To stabilize gradients and speeds up convergence during training, an "Add & Norm" layer was added after the attention layer and feed-forward network to perform residual connections and layer normalization. Multiple encoder layers can be stacked to carry out iterative attention operations. The encoder ultimately outputs a sequence that contains the most relevant historical information at each component.

To obtain the hidden representation of the input sequence, we use only the last component from this output sequence, as it contains the most relevant historical information from the entire input sequence. Finally, we map the hidden representation to stresses via a fully connected network. Overall, our framework utilizes the Transformer encoder to effectively learn the hidden representation of input information while leveraging the benefits of the self-attention mechanism.

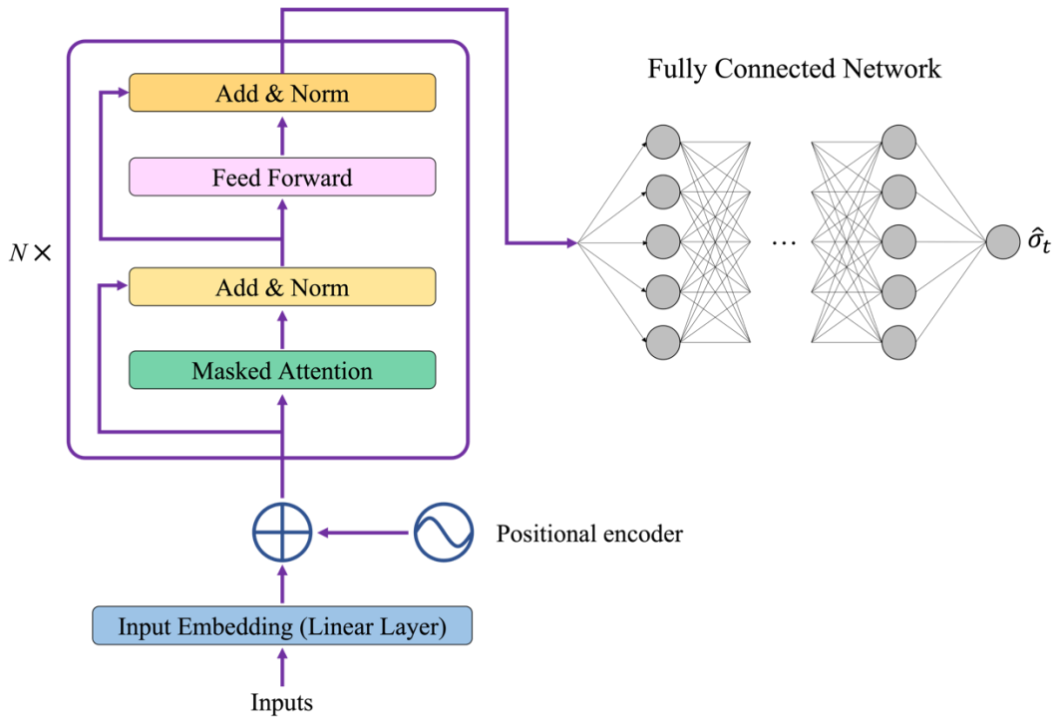

*Fig. 7 Transformer-encoder based architecture to model strain-stress relation. The transformer-encoder replaces the recurrent neural network for modeling sequence data by incorporating positional information through a positional encoder and using a masked attention mechanism to capture important history information. The output of the transformer encoder is a sequence, from which only the last component is utilized as the hidden representation of input information. The hidden representation is then mapped to stresses via a fully connected network.*

## 2.6 Datasets and model training/validation/testing

We generate two datasets to test the capability of various encoder-decoder architectures. The first dataset was obtained from uniaxial tensile tests of aluminum sheet samples. Tensile specimen design was based on previous uniaxial tensile specimen design with a gauge length-to-width ratio greater than 4 [39–41]. The specimen design and experimental setup are shown in Fig. 8.

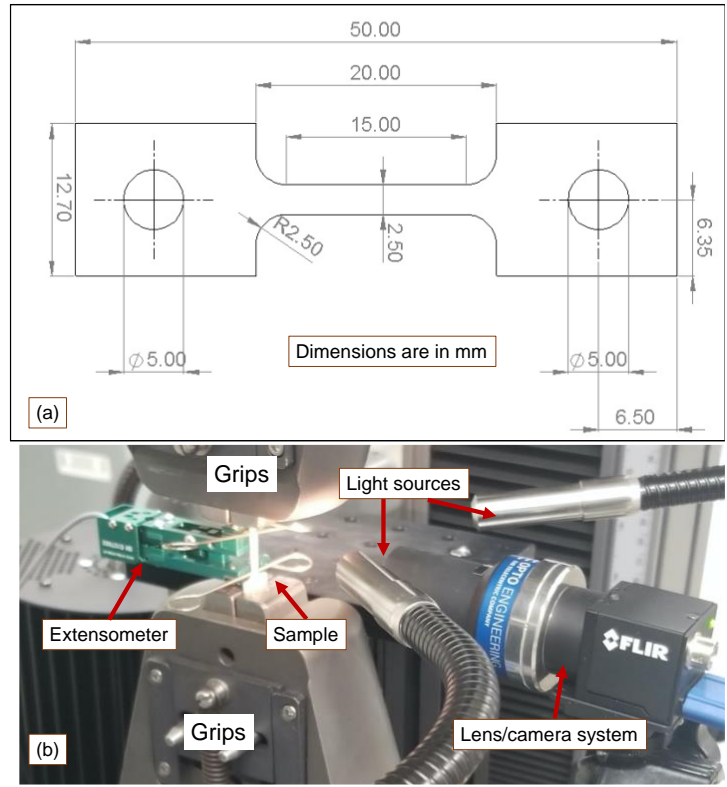

*Fig. 8 Experimental tensile tests (a) specimen geometry and (b) experimental setup.*

The test procedure followed the American Society for Testing and Materials (ASTM) E8/8M-21 standard [42] for tensile testing of metallic materials. While optical metrology data was available, the strain data was collected using a pre-calibrated Epsilon extensometer (3442-015M-050M-ST SN#E107682), and the load was measured using a calibrated 5kN Instron load cell. Both monotonically increasing and multiple loading-unloading tests, as shown in Fig. 9, were performed for the constitutive relation determination. Basic mechanical properties are listed in Table 1.

*Table 1 Unloading Basic mechanical properties of the aluminum sheet samples derived from experimental mechanical tests.*

| Test type | Yield stress (MPa) | Ultimate tensile stress (MPa) | Total elongation (%) |
|---|---|---|---|
| Monotonically increasing load | 253.7 | 319.3 | 16.5 |
| Multiple loading-unloading | 251.0[a] | 320.0 | 17.4 |

[a] based on the first loading portion

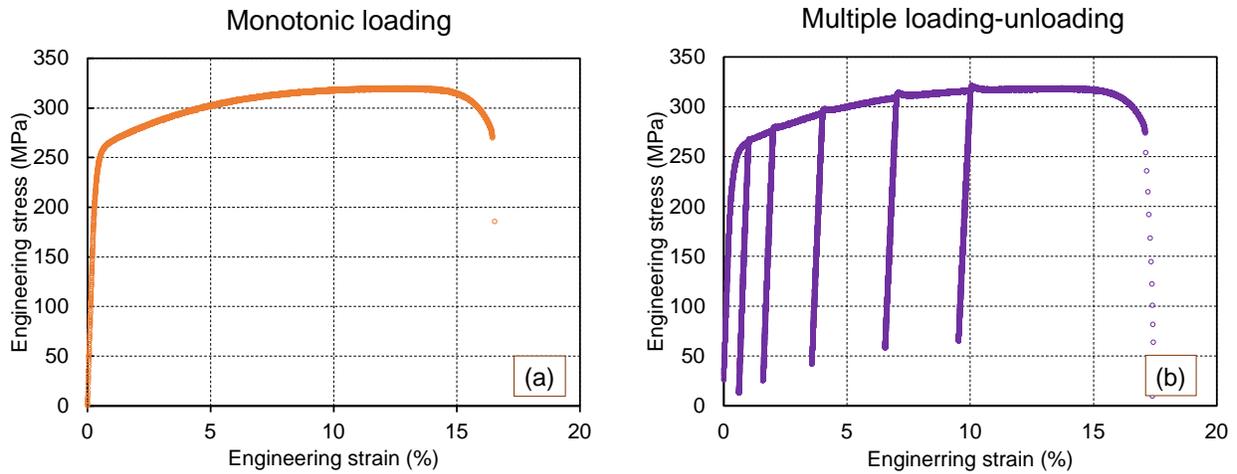

*Fig. 9 Engineering stress and strain curves of (a) monotonically increasing and (b) multiple loading-unloading test specimens.*

By applying the power law constitutive relationship to the multiple loading-unloading data (see Table 1), the strain hardening coefficient ($n$) and the constant ($K$) were calculated as 0.066 and 366 MPa, respectively. The interruption strains for unloading are shown in

Table 2.

*Table 2  Plastic strains and stresses at interruptions*

| Plastic strain (-) | Stress (MPa) |
|---|---|
| 0.006 | 263 |
| 0.0148 | 275 |
| 0.0345 | 289 |
| 0.0642 | 307 |
| 0.094 | 315 |

The use of multiple unloading/reloading paths also made the learning task more challenging. The entire dataset consists of ~8500 stress-strain pairs with a total strain up to ~18%. The loading history vector length should be chosen such that all important past points should be included to inform current decision. For example, if we want to predict the stress during an unloading process, then we should at least include the previous load-unload transition region and some strain hardening region before the load-unload transition. We used a loading history vector that contains 600 strains for each training sample (zero padding is properly added for the first zero strain).

The second dataset was generated from the Johnson-Cook model for AISI 316L steel [34], covering a wide range of loading conditions. Specifically, we included the stress-strain relations in a wide temperature range, from $10°C$ to $400°C$. For each sampled temperature, we also included the stress-strain relations at six different strain rates, i.e., $10^{-4}$ s$^{-1}$, $10^{-2}$ s$^{-1}$, $10^{0}$ s$^{-1}$, $10^{2}$ s$^{-1}$, $10^{4}$ s$^{-1}$, and $10^{6}$ s$^{-1}$. See Fig. 10 for a plot of these stress-strain relations. Each stress-strain curve has 3230 stress-strain pairs up to a total strain of 30%. We use a loading history vector of length 500 and pad zero values for the very first zero strain (such that it has the same history length).

Thus, each stress-strain curve provides 3230 sequence data, each of which contains its own loading history, temperature, and strain rate information.

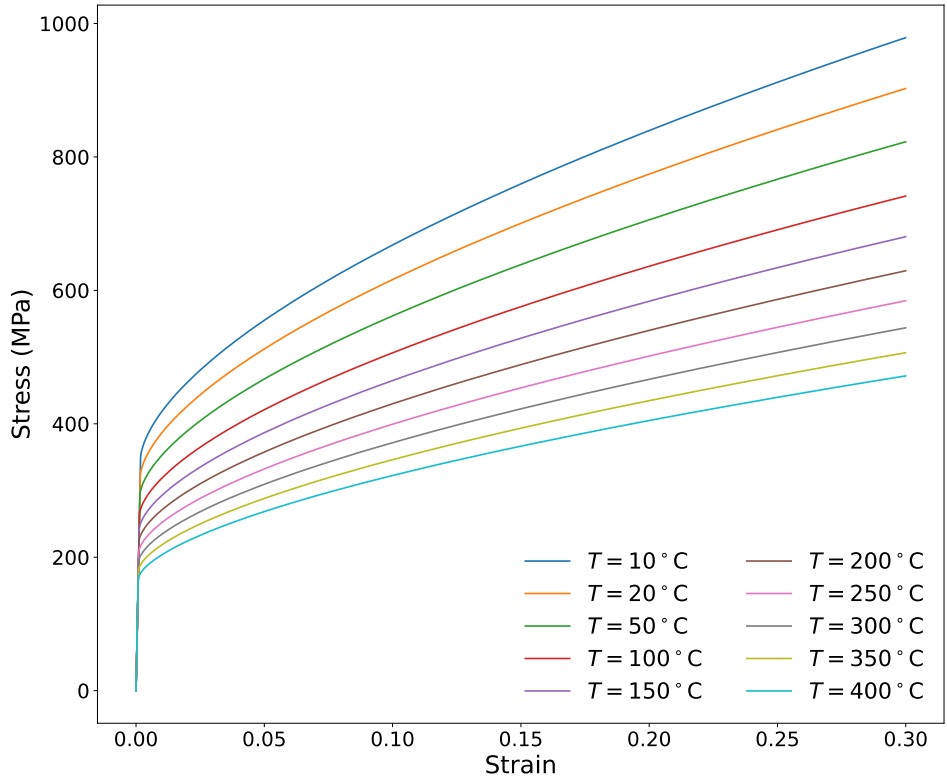

*Fig. 10 Stress-strain relations from the second datasets. These stress-strain curves were generated based on the Johnson-Cook model for AISI 316L steel[34]. Each sampled temperature also cover the stress-strain relations at six strain rates, $10^{-4}\ s^{-1}$, $10^{-2}\ s^{-1}$, $10^{0}\ s^{-1}$, $10^{2}\ s^{-1}$, $10^{4}\ s^{-1}$, and $10^{6}\ s^{-1}$.*

Both datasets were split randomly into three subsets: a training set, a validation set, and a test set, using an 8:1:1 ratio. The training set was used to train different models, while the validation set was used to select optimal model hyperparameters. After obtaining the optimal model hyperparameters, each selected model was retrained on a combined dataset consisting of the training and validation sets. Finally, the test set was used to evaluate the performance of the models. The data splitting, model training/validation/testing process is shown in Fig. 11.

To ensure that numerical scales did not influence the model training process, we performed data normalization. While strains were kept on their original scale, the temperature, strain rate, and stress quantities were standardized to have a mean of zero and a standard deviation of 1. The strain rates were log-transformed before the standardization. It's important to note that all standard scalers were obtained based on the training set and then applied to the validation and test tests, as well as any unseen data.

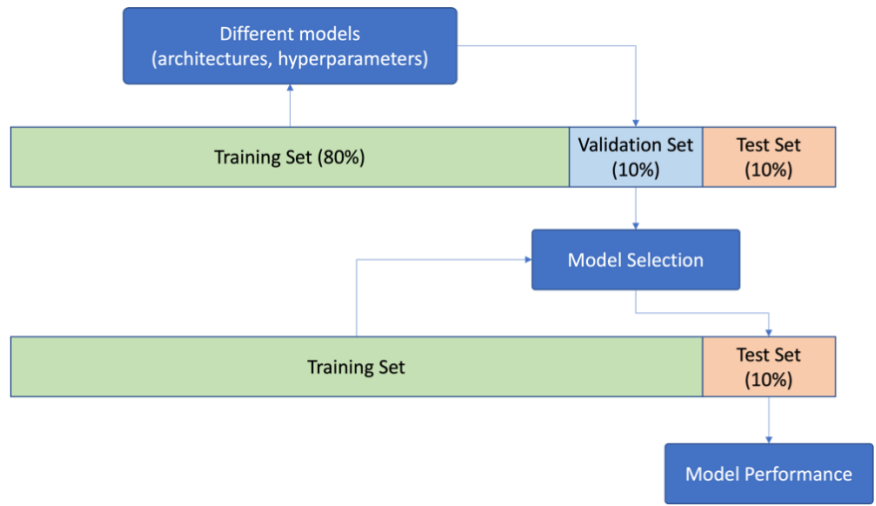

*Fig. 11 Schematic illustration of dataset splitting and model selection/testing.*

The PyTorch[43] package was used for all architecture implementation and model training. Dataset 1 models were trained for 1000 epochs, while dataset 2 models were trained for 200 epochs. An initial learning rate of $10^{-3}$ was used for all training processes, which then decays every epoch with a multiplicative factor of 0.996. We used the Adam optimizer with default hyperparameters. Model training utilized a batch size of 100 and model performance was evaluated using the root mean squared error (RMSE).

3. **Results and Discussions**

**3.1 Model valiation and testing**

To determine the optimal set of hyperparameters for each architecture, we performed model validation by training various models on the training set and evaluating their performance on the validation set. For each architecture, we started with a simple model and gradually increased its complexity to identify a robust model with a small number of model parameters as possible. Table 3 lists the important hyperparameter choices for each architecture. For the GRU encoder, we considered two critical hyperparameters: the size of the hidden state vector of the GRU cell and the number of GRU layers. In addition to these hyperparameters, for the GRU-attention encoder, we also took into account the query/value sizes, which dictate the sizes of keys/values in the attention layer. For the TCN encoder, we set the kernel size and dialation base to 2, and only focused on varying the number of channels for the causal convolution. Lastly, for the Transformer encoder, we examined the hidden vector size resulting from the attention operation and the number of encoder layers. The input embedding size and the positional encoder size were set to 32. In all cases, we used a uniform FCN decoder consisting of two hidden layers. The first hidden layer was twice the size of the hidden representation, while the second hidden layer was of the same size as the hidden representation. The output layer contained only one unit.

*Table 3 Architecture hyperparameters considered in model selection.*

|  | **GRU** | | **GRU-Attention** | | | **TCN** | **Transformer** | |
|---|---|---|---|---|---|---|---|---|
|  | hidden size | GRU layers | hidden size | GRU layers | query/value size | channel numbers | hidden size | encoder layers |
| set-1 | 5 | 3 | 5 | 3 | 5/5 | 3 | 5 | 2 |
| set-2 | 5 | 6 | 5 | 6 | 5/5 | 5 | 5 | 4 |
| set-3 | 5 | 9 | 5 | 9 | 5/5 | 8 | 10 | 2 |
| set-4 | 10 | 6 | 10 | 10 | 5/10 | 10 | 12 | 2 |

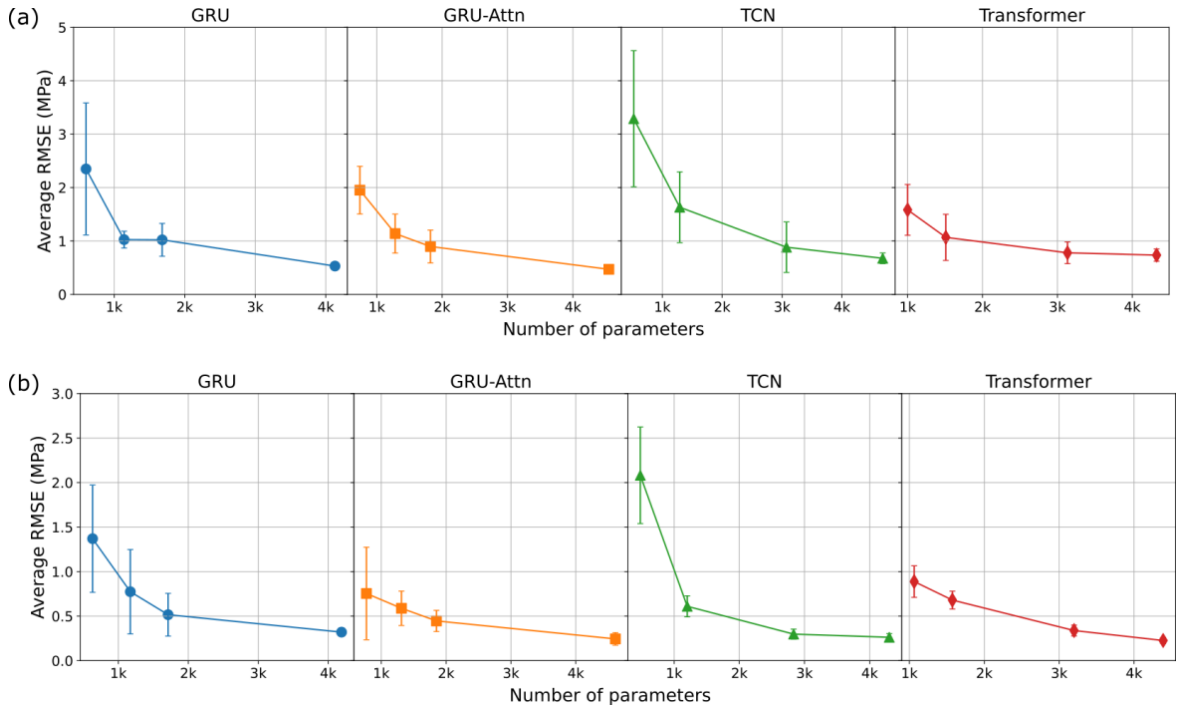

*Fig. 12 Validation results on different models. (a) Dataset 1 models. (b) Dataset 2 models.*

Fig. 12 illustrates the RMSE on the validation sets of dataset 1 and dataset 2. We varied the number of model parameters (weights and biases) by exploring different hyperparameter sets (set-1 to set-4 in Table 3). For each hyperparameter set, we trained 10 different models with different parameter initializations. The markers in Fig. 12 indicate the average RMSE over these models, and the error bars represent the standard deviations of these RMSEs. As expected, the average RMSE decreases as the model size increases. However, there are two ways to increase model size: 1) making the network deeper, as in the case of GRU and GRU-attention encoders, where the first three models (set-1 to set-3 in Table 3) have the same hidden representation size (5) but different numbers of GRU layers; and 2) using larger sizes for hidden representations, as in set-2 and set-4 for both GRU and GRU-attention encoders, set-1 to set-4 for TCN encoders, and set-1, set-3, set-4 for the Transformer encoder. Generally, deeper networks with proper connections lead to better performance, while larger hidden representations could provide more representation capability. However, increasing model complexity should be done cautiously to avoid overfitting when the training data is limited. We found that a validation RMSE smaller than 1 MPa can satisfactorily capture various stress-strain relations, including elastic deformation, strain hardening, and sharp transition during unloading-reloading cycles. Fig. 13 shows examples of models with different validation errors: the largest validation error shows relatively poor prediction on strain hardening and elastic deformation, the medium validation error predicts strain hardening well but falls short in elastic deformation prediction, and the validation error below 1 MPa captures all features in the full range deformation. Therefore, we stopped increasing the model size once the validation error converged below 1 MPa. Based on these validation results, we selected set-4 hyperparameters for each type of encoder for the testing and out-of-domain predictions.

We then performed model testing using the selected set-4 hyperparameters by retraining five different models for each type of encoder with different parameter initializations. We retrained each model on a combined dataset consisting of the original training set and validation set (see Fig. 11 for the illustration), while the original test sets were used to test these retrained models. As shown in Fig. 14, these models generally have less than 5000 parameters, but the average testing RMSE was below 0.7 MPa and 0.3 MPa for dataset 1 and dataset 2 tasks, respectively. Such testing results are comparable to the validation RMSEs shown in Fig. 12, thus we believe there is no obvious overfitting. For the dataset 1 task, we can see that GRU-based model showed the smallest testing RMSE, while GRU-attention-based model showed similar performance. The TCN-based model had the largest testing RMSE even though it had the largest model sizes, and the Transformer-based model showed intermediate testing RMSE. For the dataset 2 task, GRU-attention-based model achieved the smallest testing RMSE, showing improvements over the pure GRU-based model. The TCN-based model showed a similar performance as the GRU-attention-based model, followed by the GRU-based and Transformer-based models. The overall testing

RMSEs for dataset 2 task were smaller than that for dataset 1 task, which we believe was due to the significantly larger dataset size.

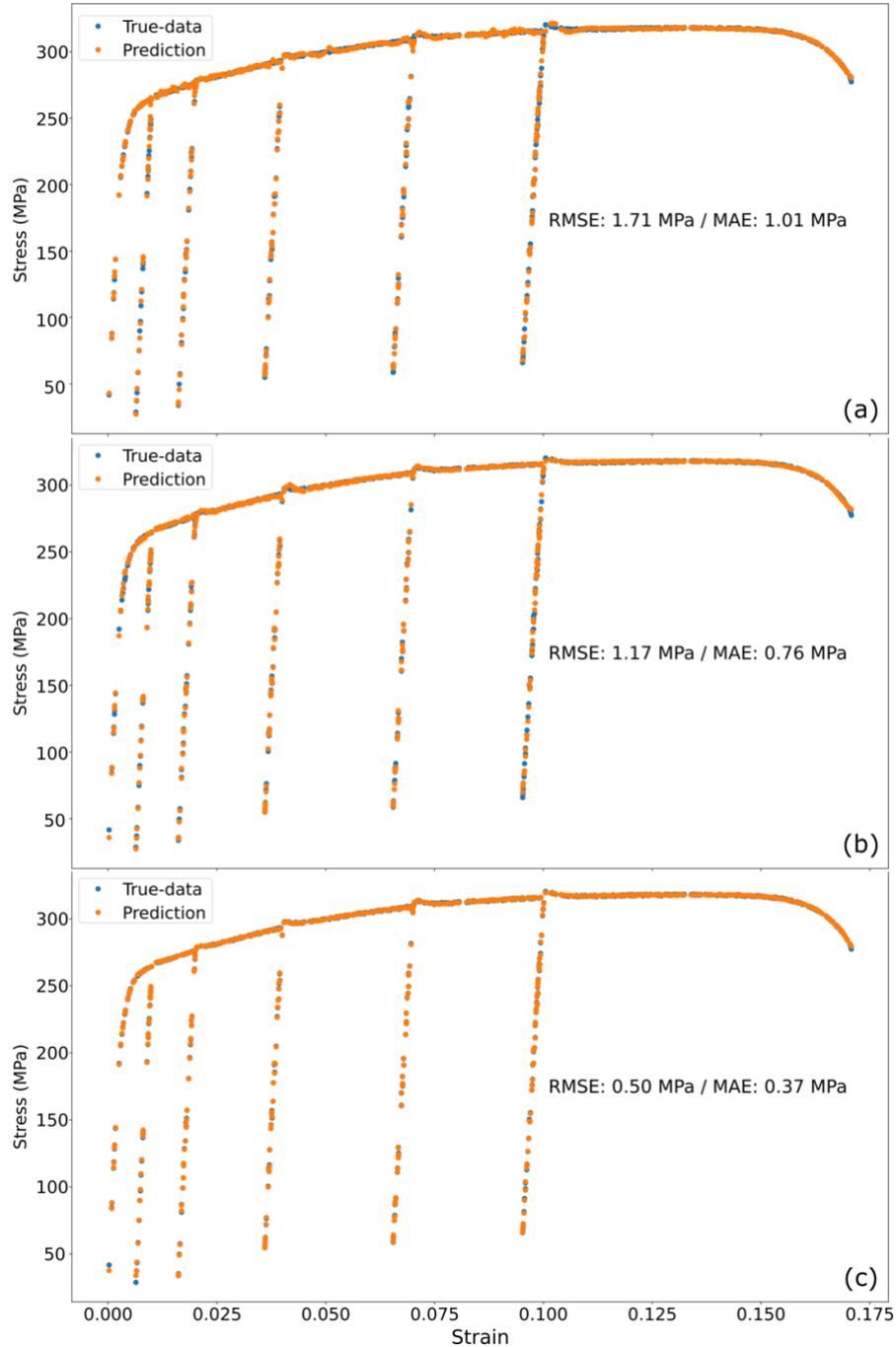

*Fig. 13 The effects of model accuracy on predicted stress-strain curves. (a-c) three GRU models showing relatively large, medium, and small RMSE on the validation set. When the validation RMSE converges below 1 MPa, all features, such as elastic deformation, strain hardening, and sharp transitions during unloading-reloading, etc., are correctly captured.*

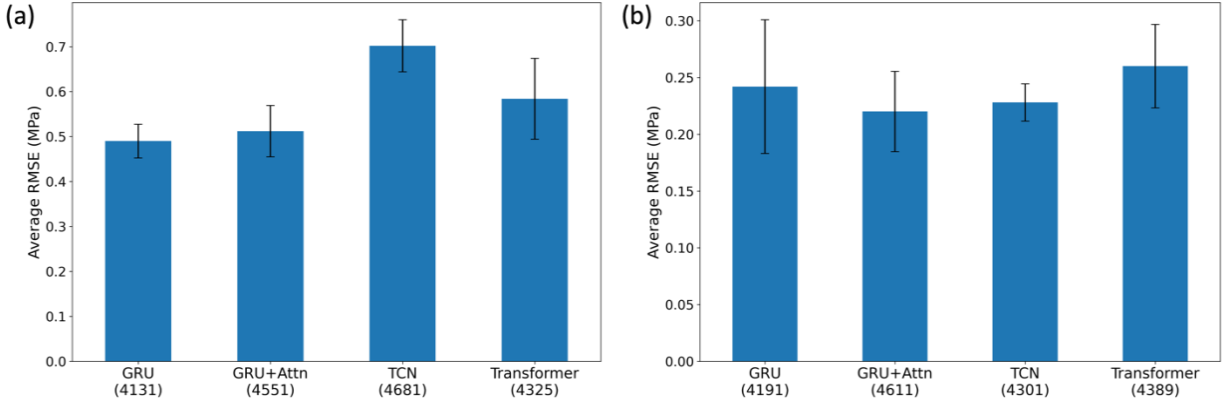

*Fig. 14 Testing results for different types of encoders using selected model hyperparameters. (a) Testing results on the test set of dataset 1. (b) Testing results on the test set of dataset 2. The average RMSE is based on five different models and the error bar represents the standard deviation of RMSEs. The number in parenthesis means the number of model parameters.*

The testing results suggest that different architectures have unique strengths in addressing different tasks, and careful choices can achieve the most robust performance. Although the overall performances of different architectures are close to each other in our demonstrated examples of tasks, two potential scenarios should be kept in mind. Firstly, the strength/weakness of different architectures may be exemplified in real-world applications that encounter datasets of different sizes and even more complex stress-strain relations. Secondly, even small reductions in validation/testing errors may lead to qualitative improvements in capturing certain features, making model selection worthwhile. Additionally, we need to consider the capability of the trained models in predicting unseen scenarios, both interpolation and extrapolation. In the following Sections, we demonstrate the applications of trained models in predicting unseen scenarios.

## 3.2 Applications of dataset 1 models in unseen loading scenarios

In this section, we present two applications of dataset 1 models to predict stress-strain behavior for unseen loading paths. The first loading path involves uniaxial loading from zero strain to final fracture strain without any unloading/reloading cycles. Although our models were trained on stress-strain curves with multiple unloading/reloading cycles, they should be able to predict a smooth strain hardening behavior in the unseen regions where unloadings were initiated, if they have learned the physical intuition of stress-strain relations.

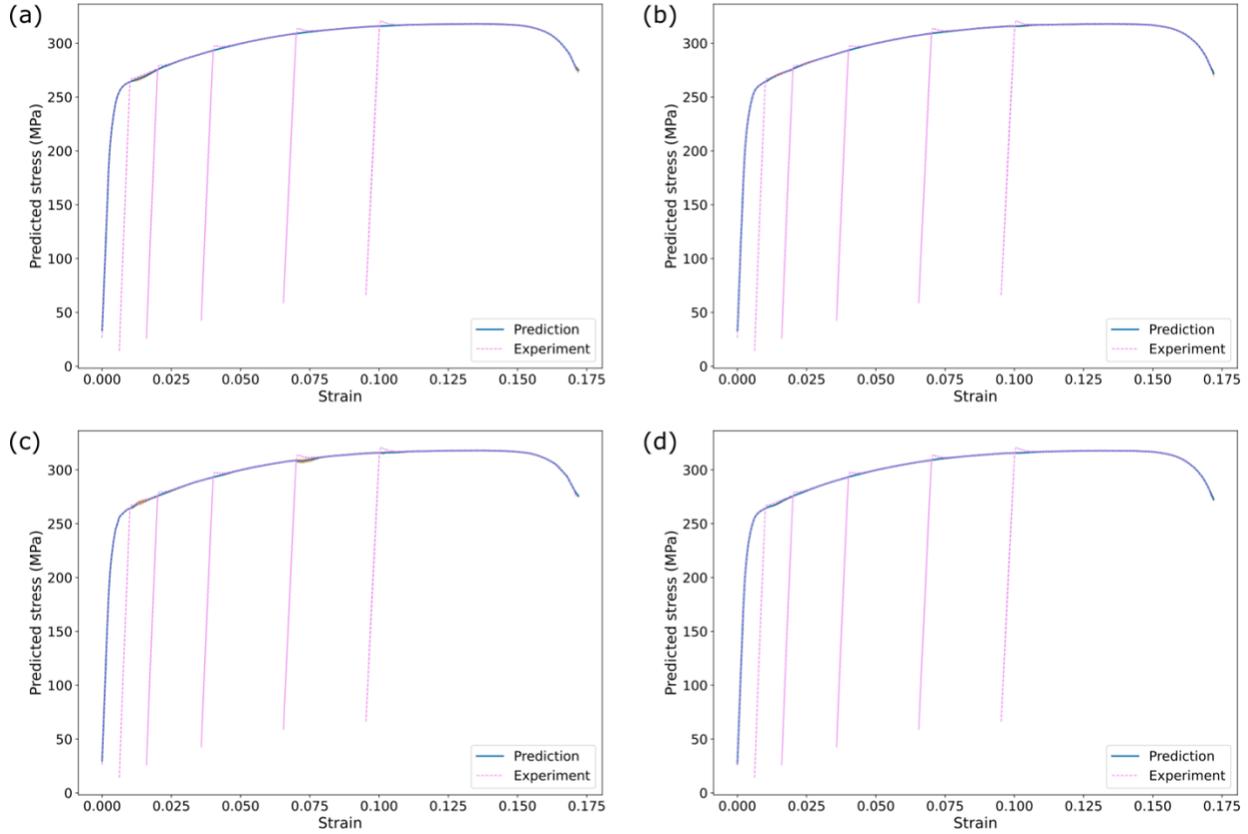

*Fig. 15 Predictions on uniaxial loading without any unloading/reloading cycles. (a) Prediction from GRU-based models. (b) Predictions from GRU-attention-based models. (c) Predictions from TCN-based models. (d) Predictions from the Transformer encoder-based models. For each prediction, the blue solid line is the average prediction from five different models and the shaded band indicates the uncertainty (based on the standard deviation of the predictions).*

Fig. 15 shows the predictions from various encoder-based models. It should be noted that the blue solid line represents the average prediction across five different models, and the shaded band indicates the uncertainty estimation based on the standard deviations of these predictions. All models were able to predict a smooth stress-strain curve, except for a few regions immediately after reloading, where predictions slightly deviated from the expected trend and showed relatively large uncertainties. The GRU-attention and Transformer encoder-based models performed better in these challenging regions. Additionally, the Transformer encoder-based models provided the most accurate predictions at the very beginning of elastic deformations. Overall, these predictions are quite satisfactory and demonstrate the models' capability of capturing the underlying physical relations.

The second loading path involves multiple unloading-reloading cycles that were not present in the training data. As demonstrated earlier, all models seem to have learned the underlying physical relations instead of just memorizing the training data pattern. Consequently, we expect the trained models to correctly predict the sharp transitions, elastic behaviors, and strain-hardening behaviors involved in unloading-reloading cycles at any point during the plastic deformation stage, including those not seen in the training dataset. To evaluate this, we introduced five unseen initiation points in the plastic deformation regime to complete unloading-reloading cycles. Fig. 16 shows various models' performance on this task. The GRU and Transformer encoder-based models both exhibit excellent predictions that adequately capture the sharp transitions, linearity of elastic deformations, and strain hardening behavior near the end of reloading. Although the GRU-attention and TCN encoder-based models showed relatively large uncertainties in certain regions, where the predicted curve was not smooth enough or deviated slightly from the expected shapes, they still captured the overall trend in a reasonable manner. These predictions on challenging unseen tasks suggest that the dataset-1 models indeed learned the underlying correlations between stress and strain for various loading scenarios.

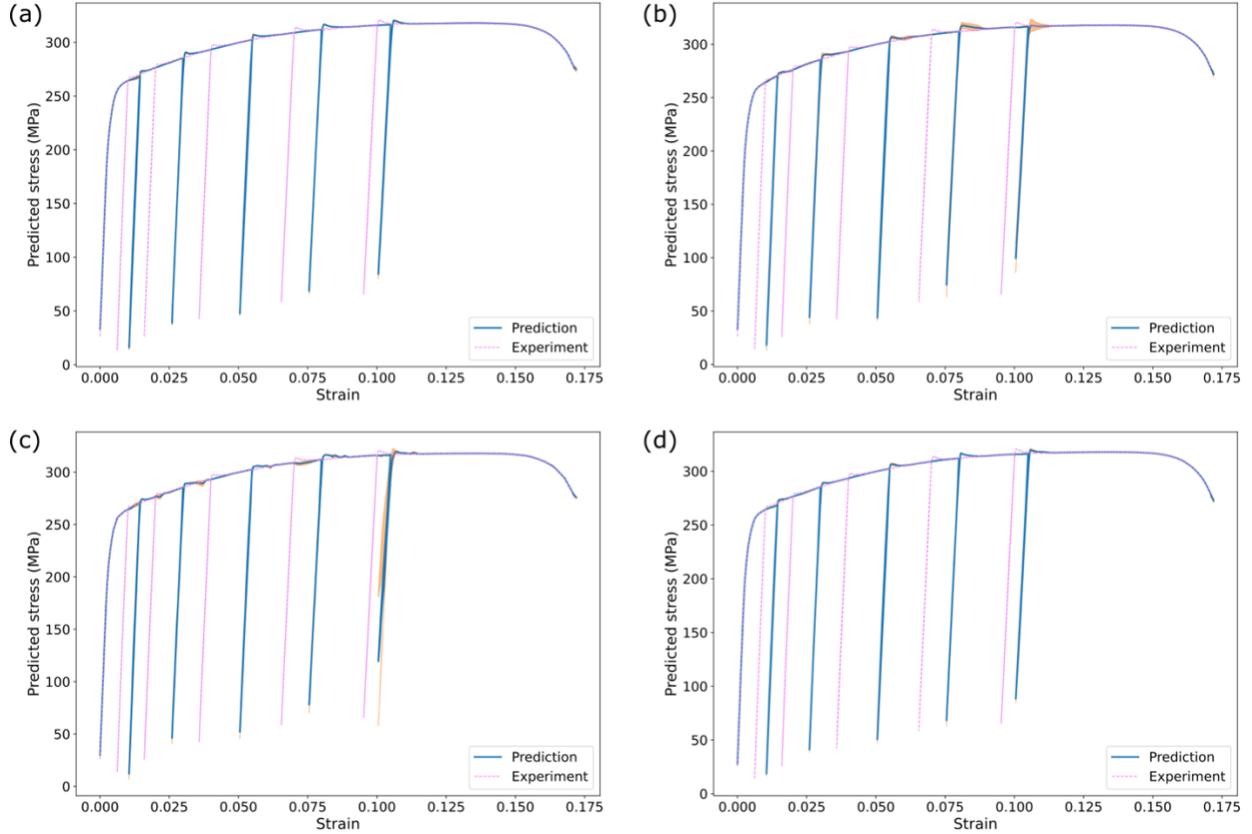

*Fig. 16 Predictions on unseen unloading/reloading cycles. (a) Prediction from GRU-based models. (b) Predictions from GRU-attention-based models. (c) Predictions from TCN-based models. (d) Predictions from the Transformer encoder-based models. For each prediction, the blue solid line is the average prediction from five different models and the shaded band indicates the uncertainty (based on the standard deviation of the predictions). Dashed lines are experimental stress-strain relations.*

### 3.3 Applications of dataset 2 models in unseen loading scenarios

In this dataset 2 task, we tested the models' performance on previously unseen loading conditions. We chose four unseen loading temperatures, including $5°C$, $125°C$, $275°C$, and $500°C$, and two unseen strain rates ($10^{-3}\ s^{-1}$ and $10^{3}\ s^{-1}$) at each temperature. Note that the temperatures $5°C$ and $500°C$ were out of the training temperature range, therefore they represent out-of-domain or extrapolation applications. Fig. 17 shows the performance of various models on these unseen loading conditions. All models performed excellently in predicting the stress-strain curves for temperatures $125°C$ and $275°C$, with their predictions (solid curves) almost overlapping with the ground truth (dashed lines, from the Johnson-Cook model) and negligible uncertainties. Although these temperatures (125°C and 275°C) were unseen during training, they were contained in the training temperature range, thus suggesting excellent interpolation capability. For the out-of-domain or extrapolation applications at 5°C and 500°C, the Transformer encoder-based model exhibited the best performance, with its predicted curves nearly overlapping with the ground truth and negligible uncertainties. The GRU, GRU-attention, and TCN encoder-based models also showed reasonable overall predictions, with slight deviations in some regions from the ground truth, indicating relatively low bias. However, these models showed relatively high uncertainties in some regions, suggesting relatively high variance. Such relatively low bias and relatively high variance indicate some overfitting, which could be mitigated by reducing the models' complexity.

Our framework is not limited to loading conditions such as temperatures and strain rates; it can also include other input information, such as materials processing conditions and materials structural/chemical information. This makes the framework useful in materials optimization processes, such as in identifying the best processing conditions in additive manufacturing. By proper model selection and achieving a bias-variance balance, the trained model can be used to predict materials mechanical behavior under unseen processing conditions and provide uncertainty estimation. This predictive capability and uncertainty estimation can be used in search strategies, such as Bayesian optimization, to effectively search optimal processing conditions. This approach can significantly reduce experimental costs, reduce human bias, and accelerate materials development.

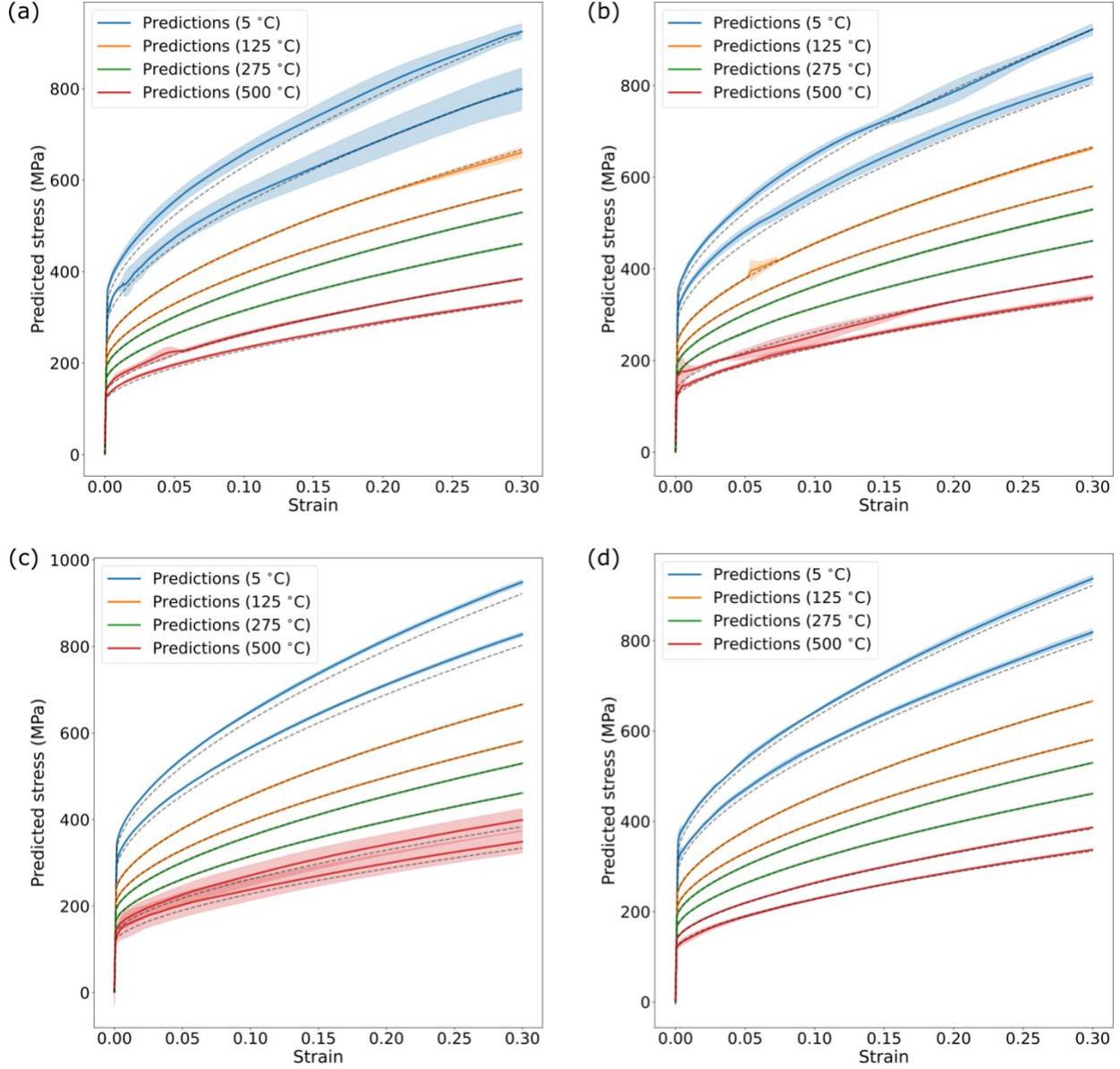

*Fig. 17 Dataset 2 models' predictions on unseen loading scenarios. (a) Predictions from GRU-based models. (b) Predictions from GRU-attention-based models. (c) Predictions from TCN-based models. (d) Predictions from the Transformer encoder-based models. Four temperatures were considered, among which $5°C$ and $500°C$ are out-of-domain temperatures. For each temperature, two strain rates, $10^{-3}$ $s^{-1}$ and $10^{3} s^{-1}$, were tested. Dashed lines are the ground truth from Johnson-Cook model. Solid lines are the average prediction across five different models, while shaded bands indicate the uncertainty based on the standard deviations of the five models' predictions.*

## 4. Summary

In this study, we developed a deep learning framework for modeling constitutive relations under various conditions, using a general encoder-decoder architecture with different encoder models such as gated recurrent unit (GRU), GRU with attention, temporal convolutional network (TCN), and Transformer. We tested and validated these models on two datasets with complex loading histories and various loading conditions. The optimal architectures with a root mean squared error converged below 1 MPa were selected for best performance in capturing various deformation behaviors. All selected architectures demonstrated excellent performance in capturing the underlying stress-strain relations, both in testing and out-of-domain scenarios. Based on the applications in this work, the Transformer encoder-based model demonstrated the best performance in both dataset tasks. Due to the universal model nature and excellent generalization ability, we expect the proposed framework to be applicable in a wide range of loading scenarios and mechanical behaviors such as fatigue loadings, hysteresis in shape memory alloys, etc. Taking advantage of the predictive capability and uncertainty estimation from deep ensembles, this framework can potentially be integrated into an active learning / Bayesian optimization [47] process for materials optimization, which can help reduce experimental cost, minimize human bias, and accelerate materials development (such as identifying optimal material processing/fabrication parameters in high throughput additive manufacturing[30,48]). Finally, we note that, in addition to robust model architectures, data consistency is also crucial for successful constitutive modeling, as different sources of stress-strain data often show uncontrolled variabilities[49].

**Acknowledgment**

This work was supported by the Laboratory Directed Research & Development Program at Idaho National Laboratory under the Department of Energy (DOE) Idaho Operations Office (an agency of the U.S. Government) Contract DE-AC07-05ID145142. This research made use of Idaho National Laboratory computing resources which are supported by the Office of Nuclear Energy of the U.S. Department of Energy and the Nuclear Science User Facilities under Contract No. DE-AC07-05ID14517. J. Li and Q-J. Li also acknowledge support by NSF CMMI-1922206.